\documentclass{article}

\usepackage{microtype}
\usepackage{graphicx}
\usepackage{subfigure}
\usepackage{booktabs} 

\usepackage{hyperref}

\usepackage[accepted]{icml2023}

\usepackage{amsmath}
\usepackage{amssymb}
\usepackage{mathtools}
\usepackage{amsthm}
\usepackage{bm}
\usepackage{physics}

\usepackage[capitalize,noabbrev]{cleveref}

\theoremstyle{plain}

\theoremstyle{definition}

\theoremstyle{remark}

\usepackage[textsize=tiny]{todonotes}

\icmltitlerunning{The galaxy-environment connection through equivariant GNNs}

\begin{document}

\twocolumn[
\icmltitle{Learning the galaxy-environment connection with graph neural networks}

\icmlsetsymbol{equal}{*}

\begin{icmlauthorlist}
\icmlauthor{John F. Wu}{1,2}
\icmlauthor{Christian Kragh Jespersen}{3}
\end{icmlauthorlist}

\icmlaffiliation{1}{Space Telescope Science Institute, 3700 San Martin Dr, Baltimore, MD 21218}
\icmlaffiliation{2}{Johns Hopkins University, 3400 N. Charles St, Baltimore, MD 21218}
\icmlaffiliation{3}{Department of Astrophysical Sciences, Princeton University, Princeton, NJ 08544, USA}

\icmlcorrespondingauthor{John F. Wu}{jowu@stsci.edu}

\icmlkeywords{Galaxy-Halo Connection, Graph Neural Networks, Galaxy Evolution, Cosmological Simulations}

\vskip 0.3in
]

\printAffiliationsAndNotice{ }

\begin{abstract}
Galaxies co-evolve with their host dark matter halos. 
Models of the galaxy-halo connection, calibrated using cosmological hydrodynamic simulations, can be used to populate dark matter halo catalogs with galaxies.
We present a new method for inferring baryonic properties from dark matter subhalo properties using message-passing graph neural networks (GNNs).
After training on subhalo catalog data from the Illustris TNG300-1 hydrodynamic simulation, our GNN can infer stellar mass from the host and neighboring subhalo positions, kinematics, masses, and maximum circular velocities.
We find that GNNs can also robustly estimate stellar mass from subhalo properties in $2d$ projection.
While other methods typically model the galaxy-halo connection in isolation, our GNN incorporates information from galaxy environments, leading to more accurate stellar mass inference.
\end{abstract}

\section{Introduction}
\label{sec:intro}

In the current $\Lambda$CDM paradigm of hierarchical galaxy formation, the galaxy-halo connection is crucial for understanding how galaxies form and evolve, and for constraining the small-scale clustering of matter \citep{SomervilleDave2015, 2018ARA&A..56..435W, Vogelsberger20_GalaxyFormation_review}. 
Techniques for modeling the co-evolution of galaxies and dark matter range from simple, non-parametric approaches to full-physics magnetohydrodynamic simulations which require $>10^8$ CPU hours of computation \citep[e.g.,][]{ValeOstriker04_AbundanceMatching, Pillepich18_mainIllustrisTNG}. 
Detailed simulations contribute important insights into galaxy formation, but due to their complexity and heavy computational costs, they are hard to analyze and cannot be performed for cosmologically significant volumes. 
Machine learning (ML) is a natural option for making progress on both of these problems.

We present an equivariant Graph Neural Network (GNN), which takes as its input a graph composed of halos linked on a linking scale of 5~Mpc, and predicts baryonic properties. 
The GNN incorporates the effects of a galaxy's environment, thereby improving the prediction of its baryonic properties compared to traditional methods. 
We are also able to train a network on the Illustris TNG300-1 box in 10 minutes on a single NVIDIA A10G GPU; inference takes one second. 
In this work, we focus on estimating stellar mass from a catalog of subhalo positions, velocities, $M_{\rm halo}$, and $V_{\rm max}$.

\section{Related work}
\label{sec:related}

The connection between galaxies and their dark matter halos has been characterized via abundance matching or halo occupation distribution (HOD) models of central halos \citep{2002ApJ...575..587B, 2002ApJ...568...52W}, conditional luminosity or mass functions \citep{2003MNRAS.339.1057Y,2010ApJ...710..903M}, subhalo abundance matching \citep[][]{2004ApJ...609...35K,ValeOstriker04_AbundanceMatching,2006ApJ...647..201C}, and empirical models of the galaxy-halo connection \citep[e.g.,][]{2013ApJ...771...30R,2019MNRAS.488.3143B}.
Several works have also attempted to perform abundance matching or paint baryons (i.e., stars) onto dark matter maps by using classical machine learning algorithms \citep[e.g.,][]{2016MNRAS.457.1162K, 2018MNRAS.478.3410A,2019MNRAS.490.2367C} and/or neural networks \citep[e.g.,][]{2019arXiv190205965Z, 2021MNRAS.507.2115M,2022ApJ...941..132M}.

In general, these previous methods treat halo/galaxy systems as unrelated entities with no formation history.  
To rectify this, \citet{2022ApJ...935...30V} construct mathematical graphs to represent group halos, and train a GNN to learn the central halo mass, which was later applied to estimate the halo masses of local Group galaxies \citep{2021arXiv211114874V}. 
GNNs have also been successfully used to model the dependence of galaxy properties on merger history \citep[e.g.,][]{2022ApJ...941....7J,2022arXiv220702786T}, and generate synthetic galaxy catalogs \citep{2022arXiv221205596J}.

In cosmology, several works have already demonstrated the representational power of GNNs, and have used it for simulation-based inference (likelihood-free inference).
\citet{2022ApJ...937..115V} employ GNNs to infer the cosmological parameters $\Omega_m$ and $\sigma_8$, using $3d$ galaxy positions  and stellar properties from the CAMELS simulation suite \citep{2021ApJ...915...71V}.
\citet{2022OJAp....5E..18M} show that GNNs can optimally extract and compress catalog data for cosmological parameter inference.
\citet{2023ApJ...944...27S} and \citet{deSanti23_camels} train GNNs to infer cosmological parameters from dark matter-only simulations, and then validate their robustness on other $N$-body and hydrodynamic simulations.

\section{Cosmic graphs}
\label{sec:method}

\subsection{Simulation data}

We use \texttt{SUBFIND} $z=0$ subhalo catalogs \citep{2001MNRAS.328..726S} derived from the Illustris TNG300-1 hydrodynamic simulation \citep{2019ComAC...6....2N,2019MNRAS.490.3196P}.
We split the full cosmological box into $6^3 = 216$ subvolumes in order to fit into 16\,GB of memory, such that each subvolume is about $(50~{\rm Mpc})^3$.
For consistency with the TNG simulations, we adopt the \citet{Planck2015} cosmology and set $H_0 = 67.74~{\rm km~s^{-1}~Mpc^{-1}}$.

\begin{figure}
    \vspace{-0.75em}
    \hspace{-1.3em}
    \includegraphics[width=0.5\textwidth]{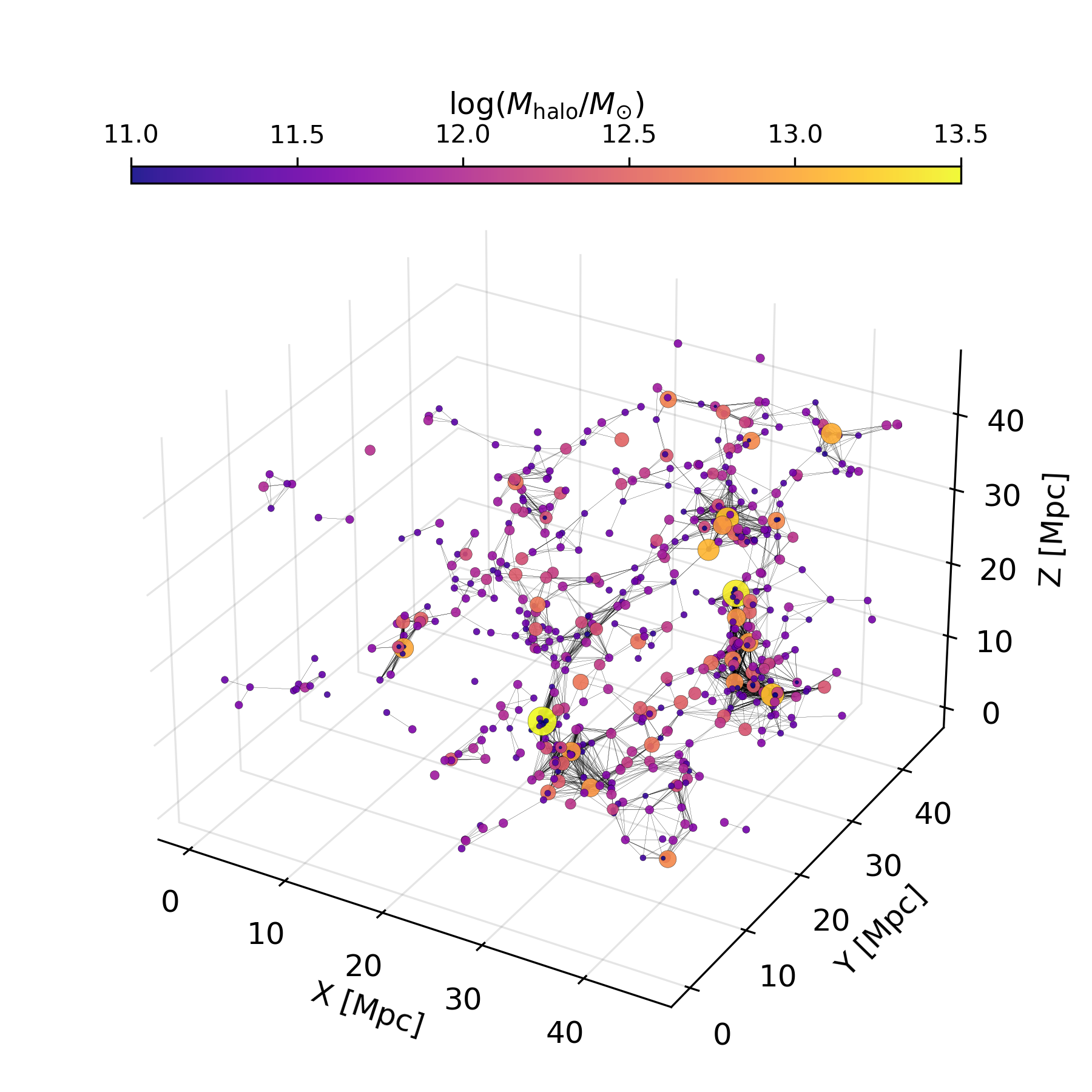}
    \vspace{-1.75em}\centering
    \caption{Cosmic graph of a TNG300 subvolume spanning approximately $45$~Mpc. Subhalos live on nodes and are colored by their logarithmic subhalo mass. Edges are formed between pairs of subhalos separated by less than the linking length of 5~Mpc.
    \label{fig:graph}
    \vspace{-1em}}
    \vspace{-1em}
\end{figure}

We select unflagged subhalos that have more than 50 star particles, $\log(M_\star / M_\odot) > 9$, and $\log(M_{\rm halo} / M_\odot )> 10$.
Due to cosmic variance, some subvolumes only have a few hundred subhalos, while others have thousands.
In Figure~\ref{fig:graph}, we show an example of a typical subvolume.

\subsection{Equivariant graph neural networks}

We construct a mathematical graph for each TNG300 subvolume, such as the one depicted in Figure~\ref{fig:graph}.
We designate $\mathcal V_i=(\vb* x_i, \vb* v_i, M_{{\rm halo},i}, V_{{\rm max}, i})$ as the eight node features. 
Subhalos within a linking length of $L = 5$~Mpc are connected with edges. Subvolumes are padded by $2.5~$Mpc on each side, such that subvolumes do not share connections that would be relevant for the linking length.
We allow nodes to be connected to themselves (i.e., self-loops).
On each edge $\mathcal E_{ij}$, we compute three features: the squared Euclidean distance $d_{ij} \equiv ||\vb* x_i - \vb* x_j||$, the inner product between unit vectors ${\vb* e}_i \cdot {\vb* e}_j $, and the inner product between unit vectors  ${\vb* e}_i \cdot {\vb* e}_{i-j}$, where unit vectors ${\vb* e}_i \equiv (\vb* x_i - \bar{\vb* x} ) / ||\vb* x_i - \bar{\vb* x}||$) are defined using positions $\vb* x_i$ relative to the centroid of the point cloud distribution $\bar{\vb* x}$, and ${\vb* e}_{i-j}$ is the unit vector in the direction of $\vb*x_i - \vb*x_j$.

\begin{figure*}[h]
    \centering
    \includegraphics[width=\textwidth]{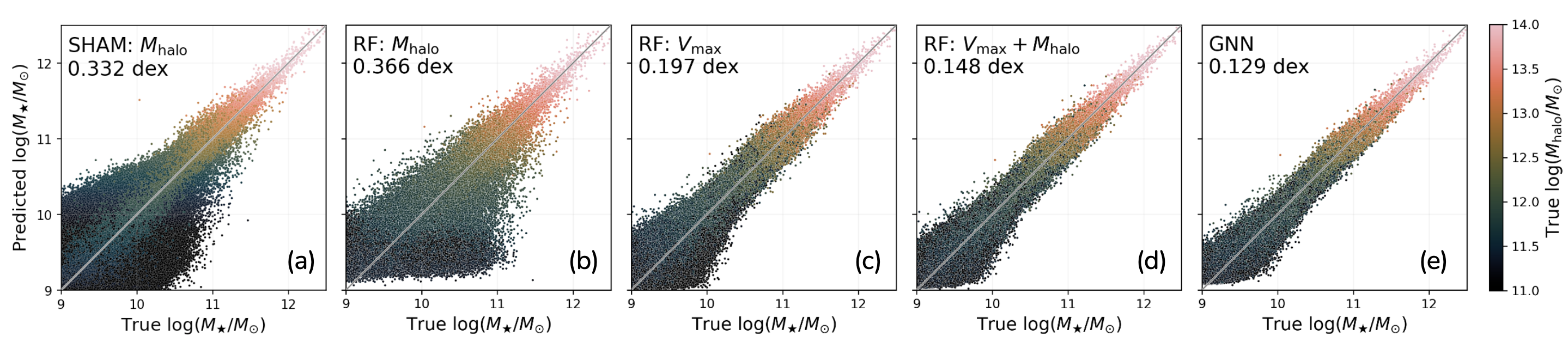}
    \vspace{-1.5em}
    \caption{Predicted stellar mass versus true stellar mass for the TNG300 data. From left to right, we show results for a subhalo abundance matching (SHAM) model, three random forest (RF) models, and our $3d$ GNN trained using $\vb* x$, $\vb* v$, $M_{\rm halo}$, and $V_{\rm max}$. We also report the scatter in the reconstructed $\log (M_\star/M_\odot)$. \label{fig:result}
    \vspace{-1em}
    }
\end{figure*}

We use a message-passing GNN based on interaction networks \citep{2016arXiv161200222B,2018arXiv180601261B}, similar to the model used by \citet{2022ApJ...935...30V}.
By design, the GNN is equivariant to permutations and invariant under the $E(3)$ group action, i.e., invariant to rotations, reflections, and translations. 
For more details about equivariant GNNs, see the appendices of \citet{2021arXiv210209844G} and Sections 3.1 and 3.2 of \citet{2022ApJ...937..115V}.
We aggregate layer inputs at each node by max pooling over information from neighboring nodes.\footnote{We do not find significant improvements by using a concatenation of sum, max, mean, and variance aggregations, or by using learnable aggregation functions.}
Our GNN has one set of fully connected layers with 256 latent channels and 128 hidden channels.
We predict two quantities for each node, which correspond to the logarithmic stellar mass $y_i \equiv \log(M_{\star,i} / M_\odot)$ and the logarithmic variance, $\log \Sigma_i$ (i.e., the logarithm of the squared uncertainty on stellar mass).

\subsection{Optimization}
Our loss function is composed of two terms: the mean squared error on the logarithmic stellar mass $||\hat{\vb* y} - \vb* y||^2$, and the squared difference between the predicted and measured variance $||\hat{\vb*{\Sigma}} - (\hat{\vb* y} - \vb* y)^2||^2$.
The latter term ensures that the variance is appropriately estimated \citep[see Moment Networks, described in Section~2 of][]{2020arXiv201105991J}.
We stabilize training by taking the logarithm of each loss term before summing them.
We monitor the loss as well as the root mean squared error (RMSE) on $\log (M_\star/M_\odot)$.

We perform $k=6$-fold cross-validation. 
For each fold, we train on 180 subvolumes and validate on 36 subvolumes, such that the validation set forms a $\sim 50 \times 300 \times 300$~Mpc$^3$ subbox.
We augment the training data set by adding random noise, sampled from a normal distribution with $10^{-5}$ times the standard deviation, for each node variable.
Based on a preliminary hyperparameter search, we implement a simple optimization schedule over a total of 1000 epochs using the \texttt{AdamW} optimizer \citep{2014arXiv1412.6980K,2017arXiv171105101L} and a batch size of 36.
We begin with a learning rate of $10^{-2}$ and weight decay of $10^{-4}$, and then decrease both by a factor of $5$ at 500 epochs, and again decrease both by a factor of $5$ at 750 epochs. 
We inspect the training and validation losses to ensure that the optimization is converged and does not overfit the training data.

\section{Results}
\label{sec:results}

Overall, we find that the GNN can infer the stellar mass from subhalo properties with remarkable accuracy.
We recover the galaxy stellar mass to within RMSE~$=0.129$~dex of its simulated value by using a GNN.
The predictions are largely unbiased as a function of mass. 

\subsection{Comparisons against baseline models}

In Figure~\ref{fig:result}, we compare the performance of different models trained and cross-validated on the same TNG300 data set.
The panels show, from left to right: (a) a subhalo abundance matching (SHAM) model, (b) a random forest (RF) trained using $M_{\rm halo}$ as input, (c) a RF trained using $V_{\rm max}$, (d) a RF trained using both $M_{\rm halo}$ and $V_{\rm max}$, and (e) a GNN trained using $3d$ positions, $3d$ velocities, $M_{\rm halo}$, and $V_{\rm max}$.
In Table~\ref{tab:metrics}, we list performance metrics for various RF and GNN models, including the RMSE, mean average error (MAE), normalized median absolute deviation (NMAD),\footnote{We define NMAD$(\bm x) \equiv k \cdot  \mathrm{median}(|x - \mathrm{median}(\bm x)|))$, where $k \approx 1.4826$ ensures that the NMAD and standard deviation are equal for a normally distributed $\bm x$.} Pearson correlation coefficient ($\rho$), correlation of determination ($R^2$), bias, and outlier fraction ($>3\times$ NMAD).

The SHAM model constructs separate monotonic relationships between $M_{\rm halo}$ or $V_{\rm max}$ and $M_\star$ for centrals and satellites.
Another difference between the SHAM model and  other approaches considered here is the former's explicit treatment of subhalo centrality. 
In order to facilitate an apples-to-apples comparison, we also train an abundance matching (AM) model that does not distinguish between satellites and centrals; however the AM model performs considerably worse than the SHAM counterpart.
We note that the AM and SHAM models are trained and evaluated on the same data set, so their performance metrics may be overinflated.

We also train several RF models, which serve as reasonable proxies for AM or conditional luminosity function models \citep[][]{2019MNRAS.490.2367C}.
By comparing panels (b) and (c), we observe that $V_{\rm max}$ is more physically connected to $M_{\star}$ than $M_{\rm halo}$, in agreement with previous findings (i.e., \citealt{
2006ApJ...647..201C,2013ApJ...771...30R}; we find this to be true for the RF, AM, and SHAM models).
A RF trained on both $M_{\rm halo}$ and $V_{\rm max}$ provides an even better reconstruction (RMSE~=~0.148~dex).

Ultimately, we find that the GNN strongly outperforms all baseline models. 
While the GNN does not distinguish between centrals and satellites, it may be able to learn whether a given subhalo is a central based on surrounding subhalo properties (see Section~\ref{sec:environments}).

\begin{table*}[t]
\caption{
    Cross-validation metrics for the AM, SHAM, RF, and GNN models discussed in the text. The GNN trained using positions, velocities, $M_{\rm halo}$, and $V_{\rm max}$ achieves the best metrics (shown in bold) in nearly every category. 
    The last two rows report metrics for the $3d$ GNN model, except that only central and satellite subhalos are selected from the cross-validation set.
    We note that the AM/SHAM models are trained and evaluated on the same data set. 
    For RF and GNN models, we repeat the entire training and cross-validation experiment three times; the scatter is too small to be shown in the displayed significant figures for all columns except the bias and outlier fraction.
\vspace{-1em}
}
\label{tab:metrics}
\vskip 0.15in
\begin{center}
\begin{tabular}{lccccccc}
\toprule
\textbf{Model} & \textbf{RMSE} & \textbf{MAE} & \textbf{NMAD} & \textbf{Pearson} $\bm{\rho}$ & $\bm{R^2}$ & \textbf{Bias} & \textbf{Outlier fraction} \\
     & (dex) & (dex) & (dex) & & & ($10^{-3}$ dex) & (\%)\\
\midrule
AM - $M_{\rm halo}$  & 0.424 & 0.327 & 0.323 & 0.736 & 0.472 & 0.1 & 3.73\\
AM - $V_{\rm max}$ & 0.173 & 0.150 & 0.132 & 0.956 & 0.912 & \textbf{0.0} & 1.91\\
SHAM - $M_{\rm halo}$   & 0.332 & 0.231 & 0.235 & 0.838 & 0.677 & 0.1 & 6.20\\
SHAM - $V_{\rm max}$   & 0.151 & 0.133 & 0.115 & 0.966 & 0.933 & \textbf{0.0} & 1.75\\
RF - $M_{\rm halo}$  & 0.366 & 0.308 & 0.277 & 0.780 & 0.606 & \bf $\bm{-0.0 \pm 0.1}$ & $2.53 \pm 0.01$\\
RF - $V_{\rm max}$  & 0.197 & 0.177 & 0.152 & 0.942 & 0.886 & $-0.3 \pm 0.0$ & $1.44 \pm 0.01$\\
RF - $M_{\rm halo}+V_{\rm max}$  & 0.148 & 0.135 & 0.114 & 0.967 & 0.936 & ~~~$0.3 \pm 0.0$ & $1.31 \pm 0.00$\\
GNN ($2d$ projection) & 0.135 & 0.131 & 0.106 & 0.973 & 0.946 & $-3.9 \pm 2.2$ & $0.68 \pm 0.01$\\
\bf GNN $\bm{(3d)}$ & \bf 0.129 & \bf 0.125 & \bf 0.102 & \bf 0.975 & \bf 0.951 & ~~~$0.8 \pm 0.6$ & $\bm{0.68 \pm 0.00}$\\
\midrule
GNN $(3d)$ - centrals & 0.123 & 0.119 & 0.097 & 0.979 & 0.959 & ~~~$4.6 \pm 0.7$ & $0.67 \pm 0.01$\\
GNN $(3d)$ - satellites & 0.138 & 0.136 & 0.109 &0.968 & 0.936 & $-5.0 \pm 0.6$ & $0.58 \pm 0.01$\\
\bottomrule
\end{tabular}
\end{center}
\vskip -0.1in
\end{table*}

\subsection{Centrals versus satellites}\label{sec:sats-centrals}

Satellite dark matter halos are preferentially stripped relative to stars in a host halo's tidal field \citep{2016ApJ...833..109S}. 
In Appendix~\ref{app:smhm}, we show the stellar mass-halo mass relation for satellite and central galaxies in TNG300 (Figure~\ref{fig:smhm}).
Indeed, we observe that satellite galaxies exhibit significantly more dispersion than centrals $M_\star$--$M_{\rm halo}$ relation. 
Our $3d$ GNN is also worse at predicting $\log(M_\star/M_\odot)$ for satellites than for centrals (see bottom two rows of Table~\ref{tab:metrics}), but this is due to the inherently larger scatter in the satellite-halo relation.
We find that there is an overall negative bias for satellites and and positive bias for centrals, because the GNN must learn separate offset relations for both centrals and satellites.

\subsection{Cosmic substructure in projection} \label{sec:projection}
We also construct cosmic graphs in projection, i.e. projected coordinates $x_1$ and $x_2$, and radial velocity $v_3$, instead of the full phase space information (see Appendix~\ref{app:projection}).
This $2d$ GNN model achieves RMSE~=~$0.135$~dex scatter, which still exceeds the performance of the best RF estimator (see Table~\ref{tab:metrics}).
Because the $2d$ GNN encode projected large scale structure information, it outperforms the RF models that can only learn isolated subhalo information.

\section{Discussion}
\label{sec:discussion}

We have presented a novel method for populating dark matter subhalos with galaxy stellar masses.
Mathematical graphs combine individual halo properties and environmental parameters in an equivariant representation, resulting in robust predictions for both central and satellite galaxies.
As shown in Table~\ref{tab:metrics} and Figure~\ref{fig:result}, the cosmic graphs outperform random forests trained on $V_{\rm max}$ and $M_{\rm halo}$.
For galaxies with $\log(M_\star / M_\odot ) \geq 9$ and $\log(M_{\rm halo} / M_\odot) \geq 10$, we recover the logarithmic stellar mass to within a root mean squared error (RMSE) of $0.129~$dex.

\subsection{Inductive biases of GNNs}
We note that previous works have employed convolutional neural networks (CNNs) for painting stars onto dark matter maps \citep{2019arXiv190205965Z,2022ApJ...941..132M}. 
Unlike abundance matching models and RFs, CNNs are able to represent local spatial information. 
However, CNNs and GNNs have different inductive biases: CNNs are well-suited for representing fields discretized onto a Cartesian grid, while GNNs are well-suited for representing objects and relationships between them.
Galaxies have small sizes ($\sim$kpc) relative to their typical separations ($\sim$Mpc), and they interact with each other (and their surrounding media) through multiple physical mechanisms (e.g., gravitational attraction, tides, ram pressure, etc).
Therefore, cosmic structures naturally conform to a graphical representation, motivating our use of GNNs in this work.

\subsection{Galaxy environments} \label{sec:environments}

We note that a GNN with no edges except self-loops would essentially model the galaxy-halo connection in isolation; all environmental information is contained and passed along the edges.
However, if we remove self-loops from the GNN, then the GNN is still able to infer $\log (M_\star/M_\odot)$ to within RMSE~$\sim$~0.145~dex.
A GNN without self-loops must estimate galaxy stellar mass \textit{solely} from neighboring halo information, which demonstrates that galaxy environments are informative for modeling the galaxy-halo connection.

We find that the GNN with max-pooling aggregation function achieves 0.001~dex lower RMSE than a GNN with sum-pooling.
This result suggests that the GNN selects the largest value for some combination of $M_{\rm halo}$, $V_{\rm max}$, and distance to neighboring subhalos in order to best make predictions.
We can speculatively interpret this as evidence that the largest and most nearby subhalo is most informative to a GNN. 
The largest subhalo might dominate environmental effects (e.g. tides and ram pressure) and control a given subhalo's stellar mass. 
Meanwhile, the summed information should capture \textit{all} of the forces, and we expect it to be more robust or transferable across domains.
This interpretation requires addition testing and an exhaustive hyperparameter search over GNN architecture and optimization procedures, which we aim to do in a follow-up work.\footnote{The linking length is a particularly important hyperparameter. In our preliminary tests, we have found 5~Mpc to give good results.}

\subsection{Applications to observations}
\label{sec:applications}
The strong performance of $2d$ GNNs (\S\ref{sec:projection}) is promising for facilitating comparisons to observations beyond the Local Group, where we can only reliably measure projected positions and line-of-sight velocities rather than full phase space information.
Our method can be used to quickly estimate galaxy properties of constrained $N$-body \citep{2022MNRAS.512.5823M} and Gpc-scale $N$-body simulated volumes \citep{2018ApJS..236...43G,10.1093/mnras/stab2484} for comparison with wide-area galaxy surveys in the low-redshift Universe \citep{2021MNRAS.502.4328R,2022ApJ...933...47C,DESI-LOWZ,2022MNRAS.513..439D,2022ApJ...927..121W}.

\subsection{Limitations and caveats}

While we have shown that the GNN outperforms other methods, this demonstration does not definitively prove that GNNs are exploiting environmental information.
Indeed, we have used a linking length of 5~Mpc, but this hyperparameter may be suboptimal and should be tuned.
It is also possible that intrinsic scatter imposes a RMSE floor \citep[i.e., due to the ``butterfly effect'' in cosmological simulations][]{2019ApJ...871...21G}, although GNN results using merger trees have shown that galaxy stellar mass can be recovered to even lower scatter \citep{2022ApJ...941....7J}.
Finally, it may be that merger history is more important than environmental information, and that the clustering information learned by a GNN only incrementally improves performance relative to other approaches.

Our results will depend on choice of halo finder, i.e. if we were to use an alternative to the \texttt{SUBFIND} algorithm (e.g. \texttt{ROCKSTAR}; \citealt{2013ApJ...762..109B}).
We have not tested our results using different halo finding tools, and it is unclear whether a GNN trained using one halo finder catalog will properly generalize to another catalog produced by a different halo finder.
We also note that our results, while promising, must be tested on dark matter only simulations with halo catalogs matched to the hydrodynamic simulation catalogs before we can rely on GNNs to paint galaxies onto dark matter subhalos.

Additionally, domain adaptation will likely be needed to ensure simulated results can transfer to other simulations (e.g., while varying cosmological parameters; \citealt{2021ApJ...915...71V}) or to observations \citep[e.g,][]{10.1088/2632-2153/acca5f}.
As a preliminary test, we repeat our experiment by training on TNG300 and validating on TNG50 data, and vice versa; in both cases the results are poor ($> 0.2$~dex).
However, by training on a subset both simulations, we can recover $\log(M_\star/M_\odot)$ to $\sim 0.13$~dex for TNG300 and $\sim 0.14$~dex for TNG50 \citep{2019MNRAS.490.3234N,2019ComAC...6....2N}.
This test suggests that cross-domain applications, such as transferring GNN results from simulations to observations, will necessitate some form of domain adaptation.

\section*{Software and Data}
Our code is completely public on Github: \url{https://github.com/jwuphysics/halo-gnns/tree/halos-to-stars}.
We have used the following software and tools:
\texttt{numpy} \citep{numpy}, 
\texttt{matplotlib} \citep{matplotlib}, 
\texttt{pandas} \citep{pandas}, 
\texttt{pytorch} \citep{Paszke2019_PyTorch}, 
and \texttt{pytorch-geometric} \citep{Fey_Fast_Graph_Representation_2019}.

We only use public simulation data from Illustris, which can be downloaded from \url{https://www.tng-project.org/data/}.

\section*{Acknowledgments}
JFW and CKJ thank Peter Behroozi, Haley Bowden, Francisco Villaescua-Navarro, Tjitske Starkenburg, and Risa Wechsler for valuable discussions that sharpened this work.
We also thank the two anonymous reviewers who provided excellent comments and suggestions that improved this manuscript.
This research has made use of NASA’s Astrophysics Data System Bibliographic Services.
The authors are grateful to the Kavli Institute for Theoretical Physics ``Building a Physical Understanding of Galaxy Evolution with Data-driven Astronomy'' program, where this work began.
This research was supported in part by the National Science Foundation under Grant No. NSF PHY-1748958.

\bibliography{main}
\bibliographystyle{icml2023}

\appendix

\section{The stellar mass-halo mass relation for satellites and centrals}
\label{app:smhm}

\begin{figure}
    \hspace{-1em}\includegraphics[width=0.95\columnwidth]{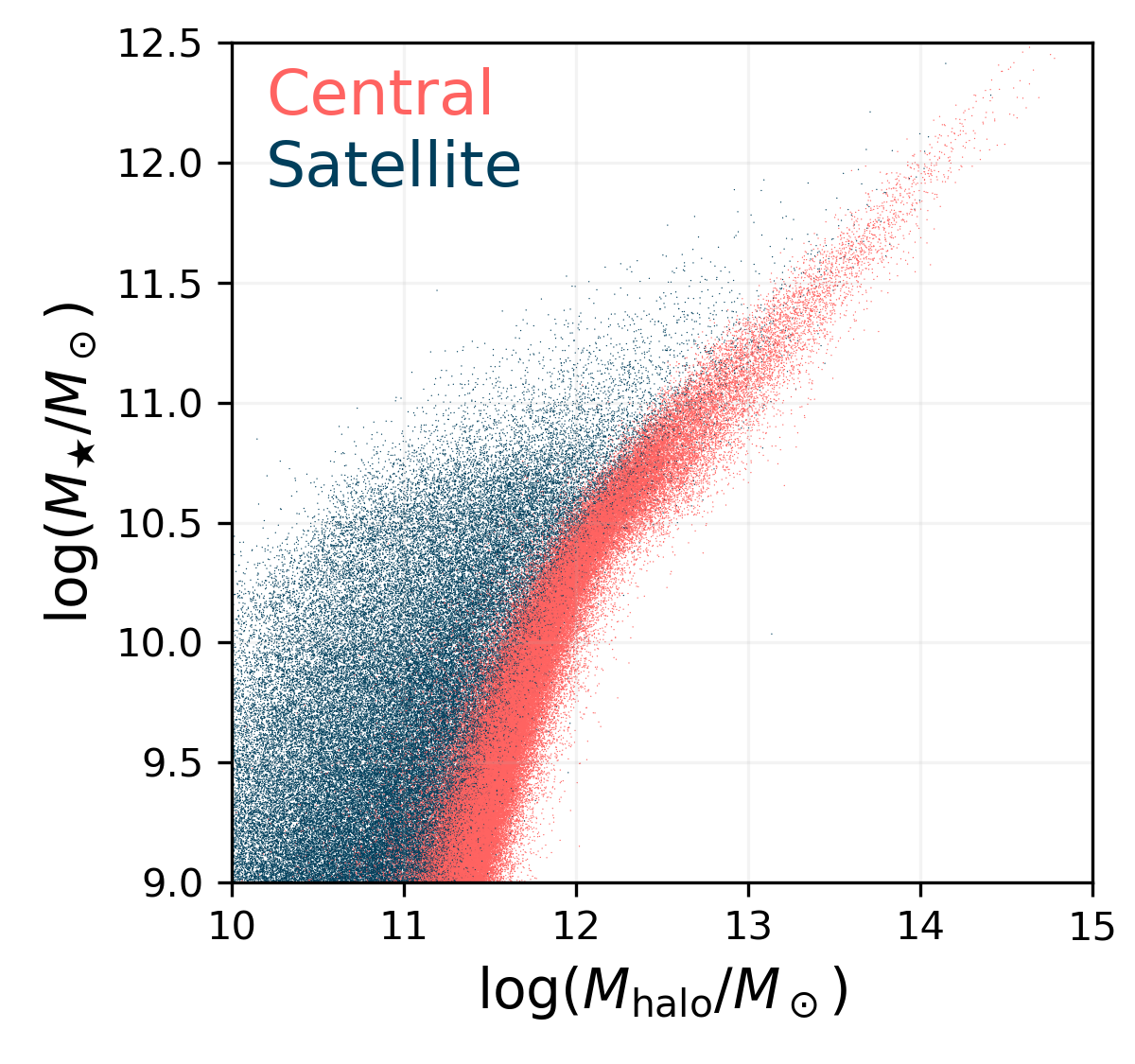}
    \vspace{-1em}\centering
    \caption{The stellar mass-halo mass relation in TNG300 for satellite and central galaxies.\vspace{-1em}
    \label{fig:smhm}
    }
\end{figure}

In Figure~\ref{fig:smhm}, we show halo masses and stellar masses for central galaxies (red) and satellites (blue) from the TNG300 \texttt{SUBFIND} catalogs.
Our GNN is able to learn the offset relationships for both central and satellite subhalos.

\section{Cosmic graphs in projected coordinates}
\label{app:projection}

\begin{figure}
    \hspace{-1em}\includegraphics[width=\columnwidth]{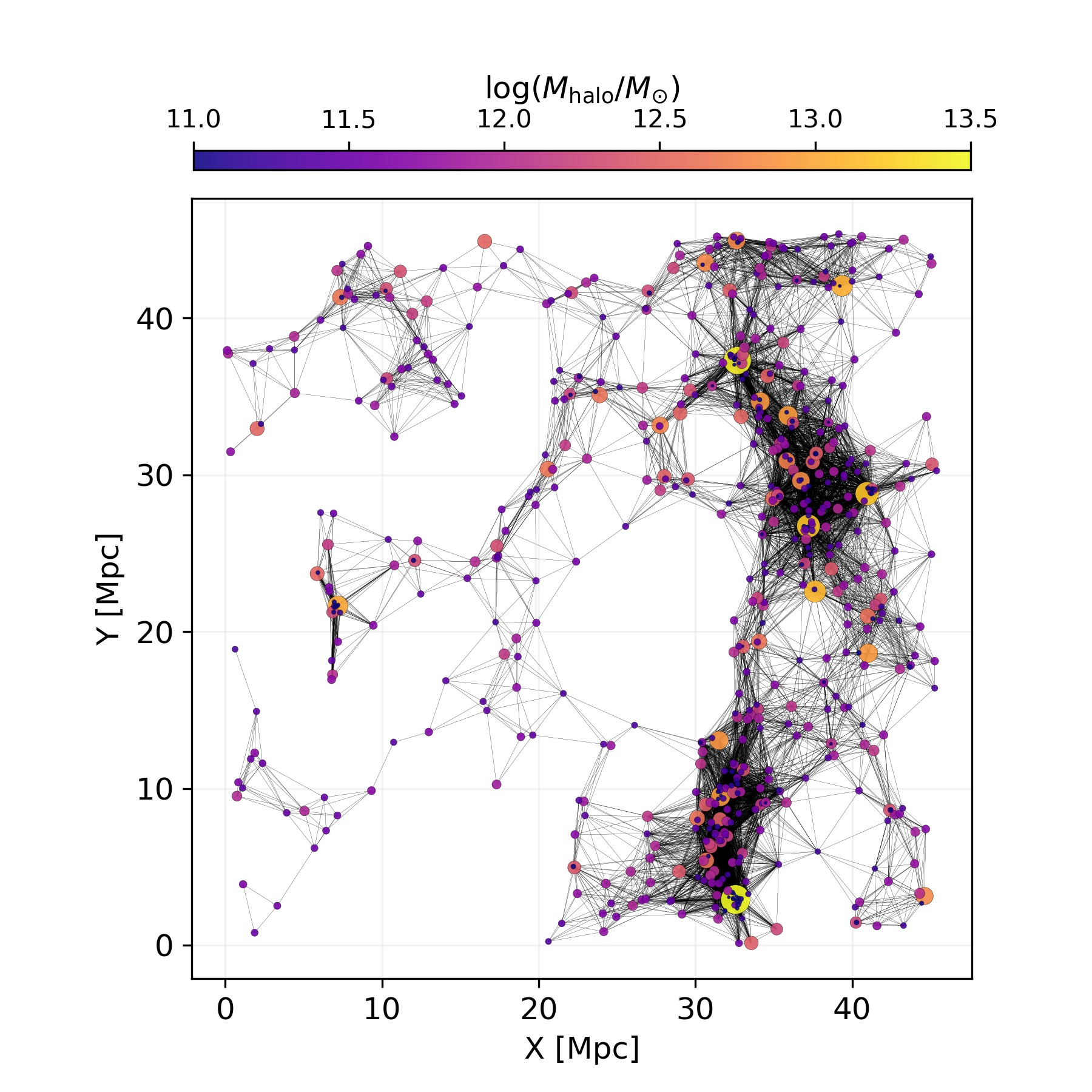}
    \vspace{-1em}\centering
    \caption{A graph of galaxies in projection, analogous to Figure~\ref{fig:graph}.
    Subhalos now connected with edges if their projected distances are less than 5~Mpc.
    \label{fig:graph-projection}
    }
\end{figure}

In \S\ref{sec:projection}, we trained a GNN to learn the galaxy-halo connection using projected positions and radial velocity, in addition to $M_{\rm halo}$ and $V_{\rm max}$.
In Figure~\ref{fig:graph-projection}, we show a projected version of the subvolume that appeared in Figure~\ref{fig:graph}.

\end{document}